\documentclass[aps,preprint,prl,superscriptaddress]{revtex4}

\usepackage{graphicx}
\usepackage{amsmath}
\usepackage{amssymb}
\usepackage{units}
\usepackage{color}

\usepackage{setspace}

\begin{document}

\title{Charge density wave melting in one-dimensional wires with femtosecond sub-gap excitation}

\author{M. Ch{\'a}vez-Cervantes}
\email{mariana.chavez-cervantes@mpsd.mpg.de}
\author{G. E. Topp}
\author{S. Aeschlimann}
\author{R. Krause}
\author{S. A. Sato}
\author{M. A. Sentef}
\author{I. Gierz}
\email{isabella.gierz@mpsd.mpg.de}
\affiliation{Max Planck Institute for the Structure and Dynamics of Matter, Center for Free Electron Laser Science, Hamburg, Germany}

\date{\today}

\begin{abstract}
Charge density waves (CDWs) are symmetry-broken ground states that commonly occur in low-dimensional metals due to strong electron-electron and/or electron-phonon coupling. The non-equilibrium carrier distribution established via photodoping with femtosecond laser pulses readily quenches these ground states and induces an ultrafast insulator-to-metal phase transition. To date, CDW melting has been mainly investigated in the single-photon and tunneling regimes, while the intermediate multi-photon regime has received little attention. Here we excite one-dimensional indium wires with a CDW gap of $\sim$300\,meV with mid-infrared pulses at $\hbar\omega=190$\,meV with MV/cm field strength and probe the transient electronic structure with time- and angle-resolved photoemission spectroscopy (tr-ARPES). We find that the CDW gap is filled on a timescale short compared to our temporal resolution of 300\,fs and that the phase transition is completed within $\sim$1\,ps. Supported by a minimal theoretical model we attribute our findings to multi-photon absorption across the CDW gap.
\end{abstract}

\maketitle

Large parallel sections on the Fermi surface of one-dimensional metals turn these systems unstable with respect to the formation of charge density waves (CDWs) where strong electron-electron and/or electron-phonon coupling gaps the electronic structure resulting in a metal-to-insulator transition below a critical temperature $T_{\text{C}}$. These CDWs and similar symmetry-broken groundstates are readily destroyed via photodoping with femtosecond laser pulses \cite{CavalleriPhysRevLett2001, IwaiPhysRevLett2003, CholletScience2005, KusarPhysRevLett2008} as photoexcitation reshapes the potential energy surface of the system and quenches the minimum associated with the symmetry-broken ground state. The associated band structure changes are commonly traced with femtosecond time- and angle-resolved photoemission spectroscopy (tr-ARPES) as a function of momentum, energy, and time \cite{PerfettiPhysRevLett2006, SchmittScience2008, RohwerNature2011, PetersenPhysRevLett2011, HellmannNatCommun2012, LiuPhysRevB2013, MathiasNatCommun2015, RettigNatCommun2015}.

Depending on photon energy, field strength, and size of the band gap, photodoping can occur via single-photon absorption, multi-photon absorption, or tunneling ionization, respectively. To date, CDW melting has been mainly investigated in the single-photon regime using pulses with photon energies of $\geq1$\,eV \cite{CavalleriPhysRevB2004, TomeljakPhysRevLett2009, HellmannPhysRevLett2010, WallPhysRevLett2012} and in the tunneling regime with strong-field Terahertz (THz) pulses \cite{RiniAppPhysLett2008, LiuNature2012, MayerPhysRevB2015, GrayPhysRevB2018}.

The recent development of mid-infrared (MIR) sources with MV/cm field strength \cite{ KaindlJOptSocAmB2000, VentalonJOptSocAmB2006} now enables us to investigate CDW dynamics in the intermediate multi-photon regime. For this purpose we excite quasi-one-dimensional indium wires that exhibit a CDW gap of $\sim$300\,meV at low temperatures with femtosecond pulses at 190\,meV photon energy and use tr-ARPES to record snapshots of the transient electronic structure. We find that the CDW gap is filled on a timescale short compared to our temporal resolution of 300\,fs indicating an ultrafast light-induced insulator-to-metal phase transition. The band structure changes are found to be complete after $\sim1$\,ps. We attribute our findings to CDW melting following multi-photon absorption across the CDW gap in good agreement with the observed intensity dependence of the absorbed energy calculated with a minimal model.

\begin{figure}
	\center
		\includegraphics[width = 1\columnwidth]{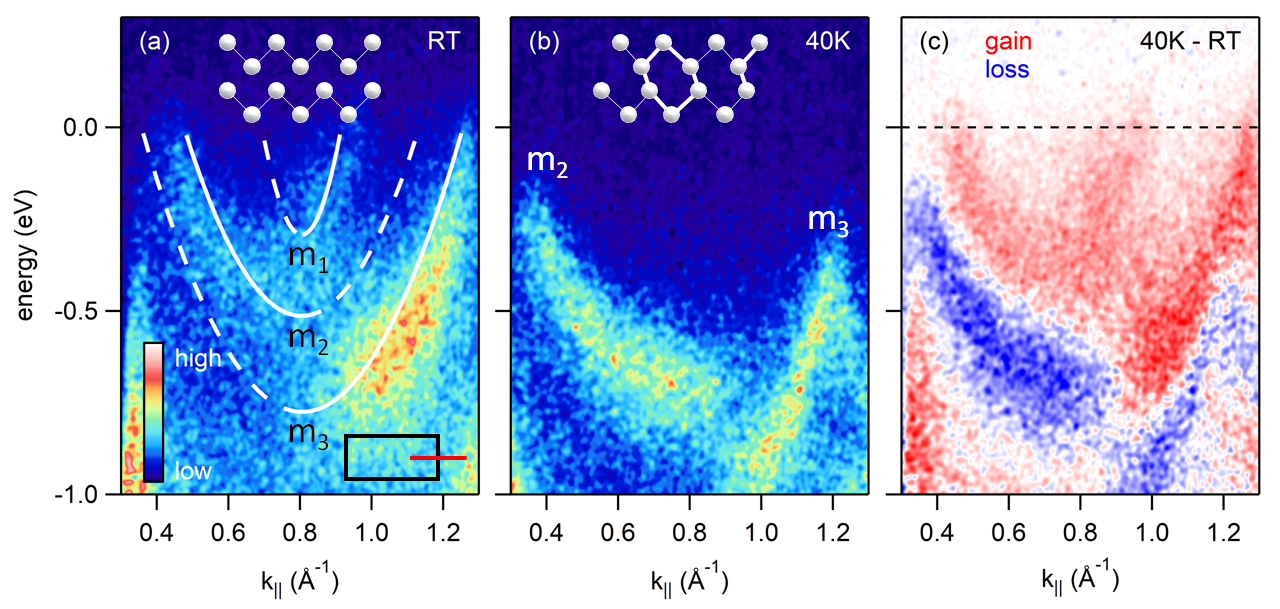}
  \caption{\textbf{Equilibrium phase transition.} Measured band structure at room temperature (a) and at T=40\,K (b). (c) Difference between panel (b) and (a). Insets in (a) and (b) show the respective structure of the Indium wires. White lines in (a) are guides to the eye based on \cite{MorikawaPhysRevB2010} that mark the position of the bands. The photemission cross section for the dashed part of each band is zero. The black box in (a) is a sketch of the ($4\times1$) Brillouin zone where the red line indicates the direction along which the band structure was measured.}
  \label{fig1}
\end{figure}

Indium atoms deposited on the $(7\times7)$ reconstruction of the (111) surface of silicon self-assemble into quasi-one-dimensional chains upon annealing at $400^{\circ}$C. At room temperature the indium atoms form pairs of zig-zag chains (see inset of Fig. \ref{fig1}a) separated by one chain of silicon atoms \cite{BunkPhysRevB1999}. The band structure has three metallic parabolic bands which cross the Fermi level at $0.41\AA^{-1}$ (m$_3$), $0.54\AA^{-1}$ (m$_2$), and $0.75\AA^{-1}$ (m$_1$) corresponding to band fillings of 0.11 (m$_1$), 0.38 (m$_2$), and 0.5 (m$_3)$ \cite{AbukawaSurfSci1995, AhnPhysRevLett2004}. The measured photoemission spectrum is shown in Fig. \ref{fig1}a. Due to photoemission matrix elements only one half of each parabola is visible \cite{MorikawaPhysRevB2010}.

Below a critical temperature of $\sim100$\,K the system undergoes a metal-to-insulator transition into a CDW ground state \cite{YeomPhysRevLett1999, AhnPhysRevLett2004, SunPhysRevB2008} with a band gap of $\sim$300\,meV \cite{ChandolaPhysRevLett2009, TanikawaPhysRevLett2004, KimPhysRevB2016} . A shear distortion that displaces the two zig-zag chains in opposite directions parallel to the wires transfers charge from m$_1$ into m$_2$. As a result both m$_2$ and m$_3$ are half-filled and susceptible to a double-band Peierls transition that is brought about by a dimerization of the outer indium atoms \cite{GonzalezNewJPhys2005, JeckelmannPhysRevB2016}. The resulting structure is that of distorted indium hexagons \cite{KumpfPhysRevLett2000} shown in the inset of Fig. \ref{fig1}b. The photoemission spectrum of the insulating phase is shown in Fig. \ref{fig1}b. 
 
In Fig. \ref{fig1}c we show the changes of the photoemission current associated with the equilibrium insulator-to-metal phase transition. We observe a pronounced loss at the position of the low-temperature dispersion, a u-shaped gain at $k_{||}<1.0\AA^{-1}$ just below the Fermi level originating from the charge transfer between m$_1$ and m$_2$, and another gain at higher momenta and larger binding energy due to the shift of m$_3$. 

A detailed description of the experimental tr-ARPES setup and the sample preparation is given in the Supplemental Material \cite{Supplementary}.

In Fig. \ref{fig2} we present tr-ARPES data taken at a base temperature of 40\,K for various pump-probe delays after photoexcitation with 300\,fs laser pulses with a photon energy of $\hbar\omega=190$\,meV and a peak electric field of 0.9\,MV/cm (see \cite{Supplementary}). Panel (a) shows the measured photocurrent for negative pump-probe delays recorded before the arrival of the pump pulse. The features are broadened with respect to Fig. \ref{fig1}b due to the presence of the pump pulse resulting in an overall energy resolution of 300\,meV for the tr-ARPES data. Panel (b) shows the pump-induced changes of the photocurrent at $t=0$\,ps, where pump and probe pulses overlap. The signal is dominated by the formation of sidebands of the unperturbed band structure due to laser-assisted photoemission (LAPE). Details are provided in the Supplemental Material \cite{Supplementary}. In Fig. \ref{fig2}c we present the pump-probe signal recorded 1\,ps after photoexcitation. The observed gain and loss features closely resemble those for the equilibrium insulator-to-metal transition in Fig. \ref{fig1}c, suggesting a transient melting of the CDW.

For a quantitative analysis of the CDW melting we integrate the pump-induced changes of the photocurrent over different areas marked by black boxes in Fig. \ref{fig2}c that highlight three main features associated with CDW melting: the in-gap spectral weight (box 1), the shift of m$_1$ below the Fermi level (box 2), and the shift of m$_3$ towards the Fermi level (box 3). The integrated photocurrent is presented in Fig. \ref{fig3} as a function of pump-probe delay. The rise time of the in-gap spectral weight in Fig. \ref{fig3}a is found to be shorter than the pulse duration of 300\,fs. From fits of the rising edge of the pump-probe signal in Figs. \ref{fig3}b and c we find that m$_1$ shifts below the Fermi level within $0.8\pm0.3$\,ps and m$_3$ shifts towards the Fermi level within $1.1\pm 0.4$\,ps. These numbers are similar albeit slightly longer compared to those obtained for photoexcitation at $\hbar\omega\geq1$\,eV in previous time-resolved electron diffraction \cite{WallPhysRevLett2012, FriggeNature2017} and tr-ARPES investigations \cite{ChavezPhysRevB2018, NicholsonArxiv2018}. For a field strength of 0.9\,MV/cm we find a long-lived transient state, indicating complete melting of the CDW \cite{WallPhysRevLett2012, FriggeNature2017}.

After providing direct experimental evidence for CDW melting in one-dimensional indium wires with strong pump pulses the photon energy of which is smaller than the CDW gap in Figs. \ref{fig2} and \ref{fig3} we now set out to unravel the underlying mechanism. Nicholson et al. \cite{NicholsonArxiv2018} previously identified the formation of zone-boundary holes as the primary driving force for the light-induced phase transition in one-dimensional indium wires.

With that in mind we can compare our findings to a related insulator-to-metal phase transition observed in VO$_2$ following photoexcitation with strong MIR and THz pulses with photon energies $\hbar\omega<E_{gap}=670$\,meV \cite{RiniAppPhysLett2008, LiuNature2012, MayerPhysRevB2015, GrayPhysRevB2018}. Rini et al. \cite{RiniAppPhysLett2008} suggested that photoabsorption at $\hbar\omega=180$\,meV, enabled by the existence of in-gap defect states in polycristalline VO$_2$ films, drives the phase transition via the generation of photoholes in the valence band of VO$_2$. In contrast to this, Mayer et al. \cite{MayerPhysRevB2015} as well as Gray et al. \cite{GrayPhysRevB2018} used strong THz transients to drive the insulator-to-metal phase transition in VO$_2$ which they attributed to tunneling ionization across the gap. 

Aside from defect-mediated sub-gap absorption and tunneling ionization, multi-photon absorption emerges as a third scenario that might provide the non-equilibrium carrier distribution required to drive the insulator-to-metal phase transition in the indium wires used in the present study. While the importance of defect-mediated in-gap states is difficult to assess in the present tr-ARPES study, we can easily distinguish between tunneling ionization and multi-photon absorption by considering the Keldysh parameter $\gamma$ that is given by the square root of the ratio between ponderomotive energy and gap size \cite{KeldyshJExpTheorPhys1965}. $\gamma>1$ and $\gamma<1$ correspond to photoexcitation via multi-photon absorption and tunneling, respectively \cite{KeldyshJExpTheorPhys1965}. From the experimental parameters we obtain $\gamma_{\text{exp}}$=1.6 (see \cite{Supplementary}) implying multi-photon absorption.

In order to support this interpretation, we simulate the interaction of the MIR pump pulse with our sample (details are provided in \cite{Supplementary}) and calculate the absorbed energy for different field strengths as a function of pump-probe delay shown in Fig. \ref{fig4}a. We find that for field strengths above $\sim$1\,MV/cm the absorbed energy exceeds the CDW condensation energy of 32\,meV at T=40\,K \cite{WippermannPhysRevLett2010} and the CDW is expected to melt. The theoretically predicted threshold field agrees well with the experimental value of $\sim0.9$\,MV/cm. The small discrepancy between the two values is attributed to uncertainties regarding the size of the gap (literature values range from $\sim$120\,meV \cite{YeomPhysRevB2002, SunPhysRevB2008} to 350\,meV \cite{ChandolaPhysRevLett2009, NicholsonArxiv2018}) and the condensation energy (literature values range from 47\,meV \cite{WippermannPhysRevLett2010} to 66\,meV \cite{KimPhysRevB2016} at T=0\,K). Further, both the CDW gap and the condensation energy are temperature dependent. Therefore, both quantities are expected to decrease as a function of time when the pump pulse hits the sample. As the model uses a fixed gap size of 300\,meV it is expected to overestimate the field required to melt the CDW.

The log-log plot of absorbed energy versus intensity (field squared) in Fig. \ref{fig4}b allows us to distinguish between multi-photon and tunneling regime as the two regimes exhibit different slopes. For fields below 1.3\,MV/cm we fit a slope of two indicating that each photoexcited electron absorbs two photons. For higher fields the absorbed energy exhibits a slower increase indicating tunneling ionization. Note that the deviation from the quadratic intensity dependence coincides with the crossover between $\gamma>1$ and $\gamma<1$ (blue-shaded are in Fig. \ref{fig4}b) and that the threshold field for CDW melting lies in the multiphoton regime in agreement with the experimental Keldysh paramter.

In summary, we have used tr-ARPES to probe the band structure changes induced by strong-field MIR excitation at photon energies smaller than the CDW gap in quasi-one-dimensional indium wires. We observe a transient melting of the CDW on a timescale that is similar to the one observed at pump photon energies $\hbar\omega\geq1$\,eV. Supported by a minimal model we attribute our results to photodoping via multi-photon absorption that presumably generates holes at the zone boundary. Our findings are relevant for ultrafast optical switching and open new pathways for MIR detection. 

We thank A. Rubio for many helpful discussions and acknowledge financial support from the German Science Foundation through the Collaborative Research Center CRC925 and the Emmy Noether program (SE 2558/2-1).

\clearpage
\pagebreak

\begin{figure}
	\center
		\includegraphics[width = 1\columnwidth]{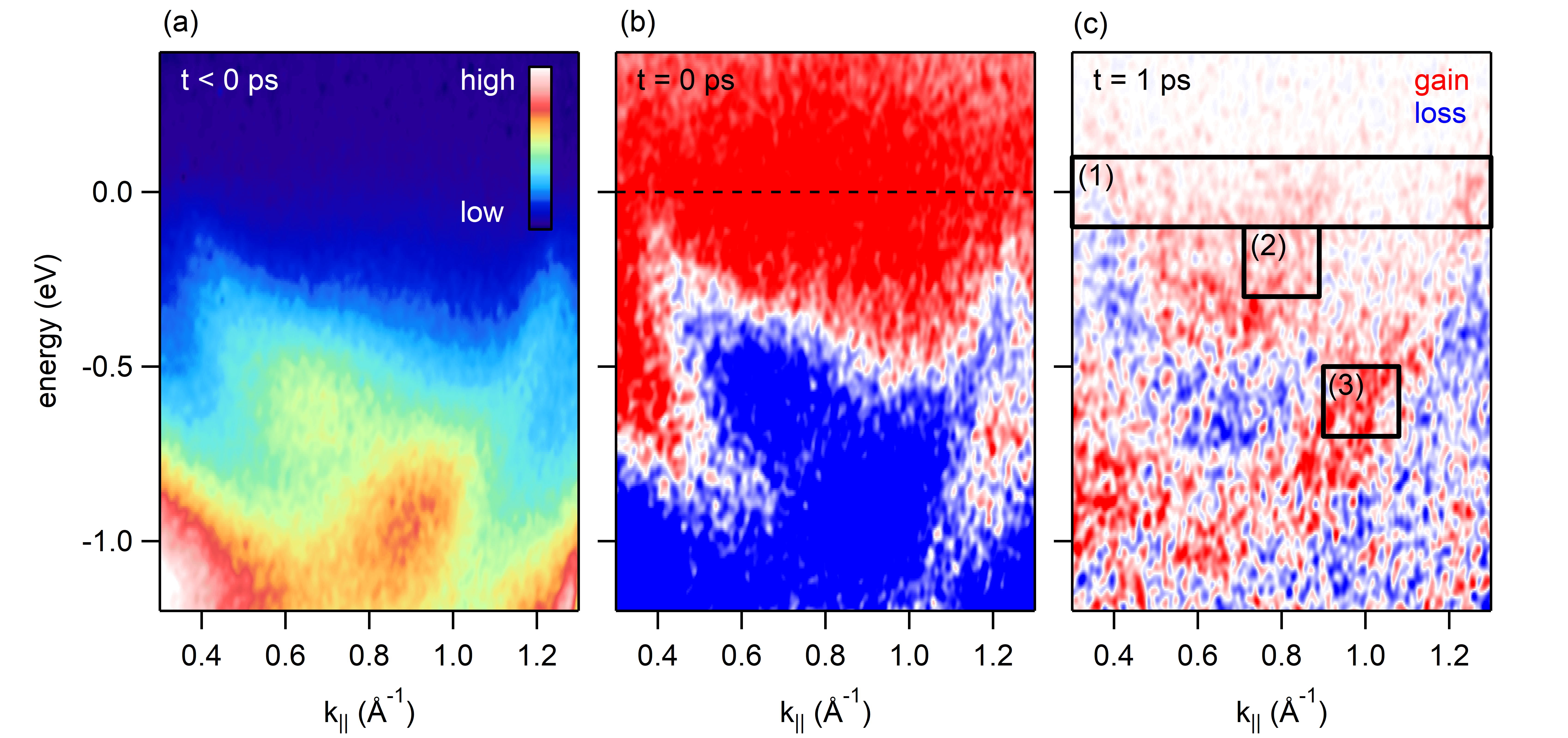}
  \caption{\textbf{Light-induced phase transition.} (a) Measured band structure at negative pump-probe delay. (b) Pump-induced changes of the photocurrent in the presence of the 300\,fs pump pulse at $\hbar\omega=190$\,meV with a peak electric field of 0.9\,MV/cm. (c) Same as (b) but for a pump-probe delay of 1\,ps. Black boxes in (c) indicate the area over which the pump-probe signal in Fig. \ref{fig3} was integrated.}
  \label{fig2}
\end{figure}

\begin{figure}
	\center
		\includegraphics[width = 1\columnwidth]{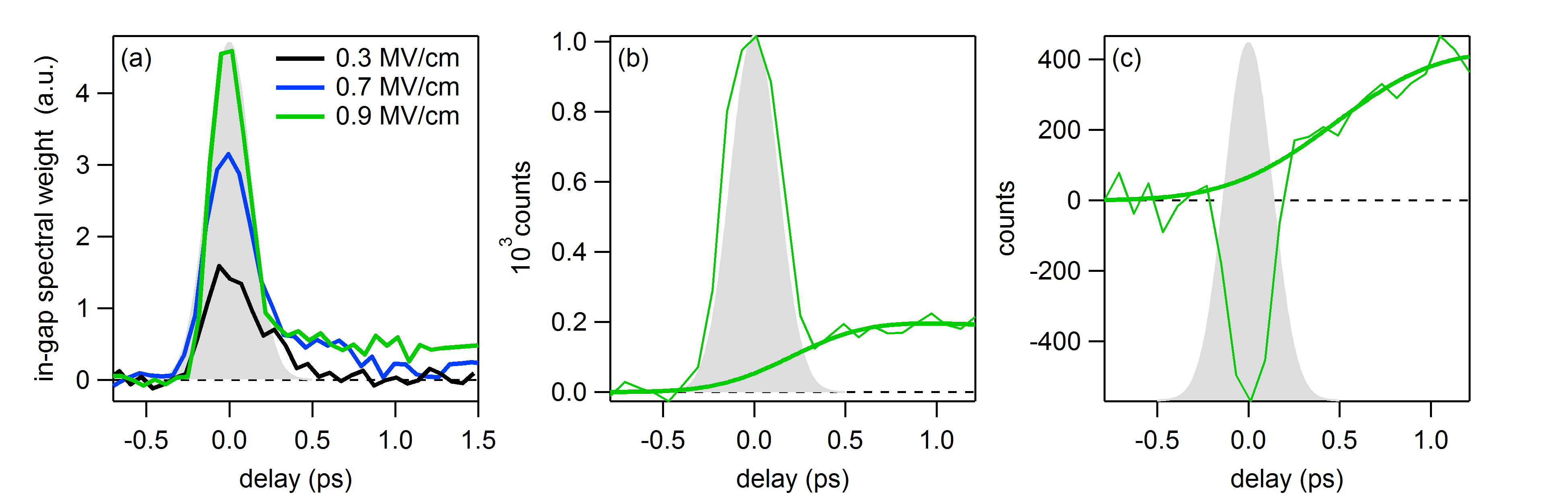}
  \caption{\textbf{CDW melting dynamics.} Pump-probe signal obtained by integrating the pump-induced changes of the photocurrent over the areas marked by the black boxes in Fig. \ref{fig2}c. The gray-shaded region is the pump pulse. (a) In-gap spectral weight as a function of pump-probe delay for different peak field strengths from box (1). The rise time of the in-gap spectral weight is shorter than the pump pulse. (b) Downshift of m$_1$ from box (2) with a rise time of $0.8\pm0.3$\,ps. (c) Upshift of m$_3$ from box (3) with a rise time of $1.1\pm0.4$\,ps. The rise time is given by the full width at half maximum of the derivative of the rising edge.}
  \label{fig3}
\end{figure}

\begin{figure}
	\center
		\includegraphics[width = 1\columnwidth]{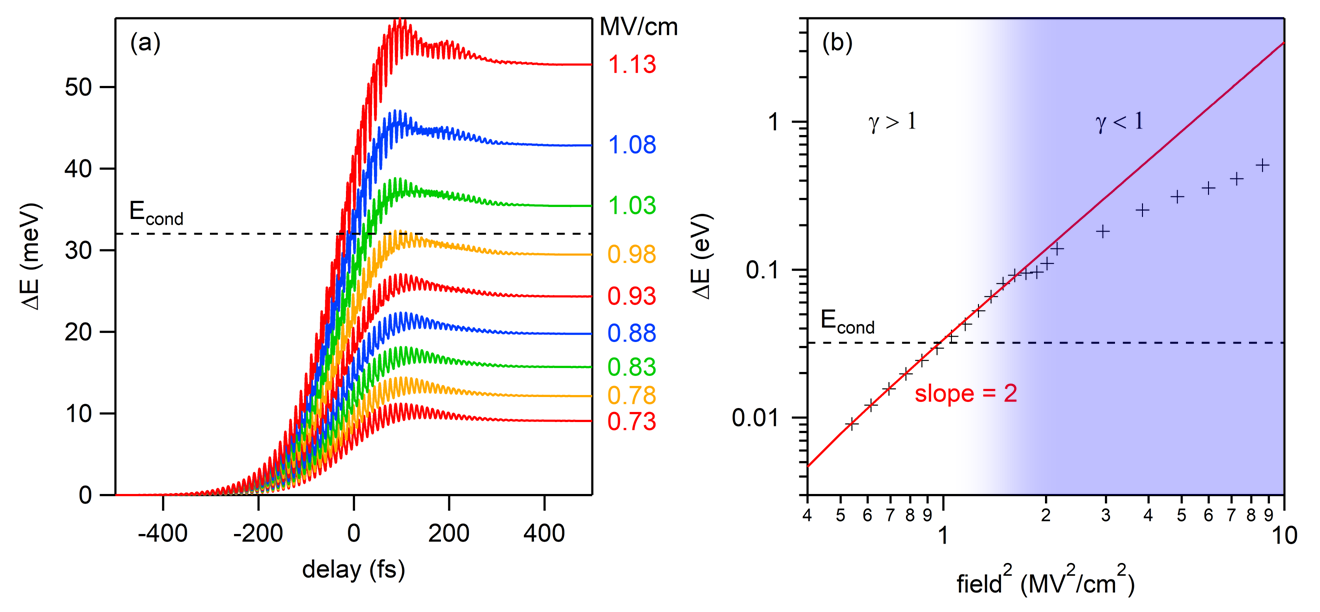}
  \caption{\textbf{Simulation.} (a) Absorbed energy as a function of time for different peak field strengths. The dashed line indicates the CDW condensation energy of 32\,meV per $(4\times2)$ unit cell. (b) Log-log plot of absorbed energy versus the square of the peak field. The red line has a slope of two as expected for two-photon absorption. White and blue areas distinguish the multi-photon $(\gamma>1)$ and the tunneling regime $(\gamma<1)$, respectively.}
  \label{fig4}
\end{figure}

\clearpage
\pagebreak

\section{Supplementary Material}

\subsection{Sample preparation}

The samples were prepared on a phosphorous-doped Si(111) wafer from CrysTec. The sample has a resistance ranging from 1 to 20\,$\Omega$cm and a $1^{\circ}$ miscut along the [$\bar1\bar1$2] direction. The small miscut angle ensures the growth of indium wires in a single domain \cite{StevensPhysRevB1993}. The wafer was annealed to 1100\,$^{\circ}$C by direct current heating until the pressure in the chamber stayed below $1\times10^{-9}$\,mbar. In order to obtain regular steps of mono-atomic height on the ($7\times7$) surface we flashed the substrate to 1260\,$^{\circ}$C and slowly cooled down to 1060\,$^{\circ}$C followed by a fast temperature decrease to 850\,$^{\circ}$C \cite{LinJApplPhys1998}. We repeated this procedure until the pressure stayed below $3\times10^{-9}$\,mbar during the 1260\,$^{\circ}$C flash. Afterwards we deposited around 10 monolayers of indium on the clean substrate at room temperature from an electron beam evaporator and annealed the sample at 400\,$^{\circ}$C for 5 minutes which produced the desired ($4\times1$) structure. All steps during sample preparation were monitored with low energy electron diffraction (LEED).

\subsection{Tr-ARPES setup}

The tr-ARPES setup is based on a Titanium:Sapphire amplifier operating at a repetition rate of 1\,kHz to generate synchronized pump and probe pulses. 1\,mJ of output power is frequency-doubled in a $\beta$ barium borate (BBO) crystal and focused into an argon jet to generate high harmonics, producing a broad spectrum of extreme ultra-violet (XUV) light. The 7$^{th}$ harmonic at $\hbar\omega_{\text{probe}}=22$\,eV is selected with a time-preserving grating monochromator \cite{FrassettoOptExpress2011} and used as a probe pulse for the tr-ARPES experiments. This photon energy is high enough to reach beyond the first Brillouin zone boundary of the In/Si(111) ($4\times1$) phase and measure the complete band structure of the system. 

Mid-infrared pump pulses are generated by overlapping the signal at 1410\,nm and idler pulses at 1796\,nm from an optical parametric amplifier on a GaSe crystal to obtain pulses at 6.6\,$\mu$m ($\hbar\omega=190$\,meV) by difference frequency generation. 

After excitation of the sample with the pump pulse the time-delayed XUV pulse ejects photoelectrons. Snapshots of the band structure are obtained by dispering the photoelectrons according to their kinetic energy and emission angle with a hemispherical analyzer and counting them with a two-dimensional detector.

The overall energy and time resolution of the tr-ARPES setup for this experiment were 300\,meV and 300\,fs, respectively.

\subsection{Calculation of the peak electric field}

We estimate the peak electric field in the surface of the sample parallel to the wires as follows. First we measure the average power $P$ and the spot size $d$ (full width at half maximum of the Gaussian beam profile) at the sample position. The pump pulse duration $\tau$ is obtained by fitting the time dependence of the LAPE signal (see next section) with a Gaussian the width of which is given by the cross correlation between pump and probe pulses. The probe pulse has a nominal duration of 100\,fs. From these parameters we calculate the peak intensity of the pump pulse via

\begin{equation}
I=\frac{4P}{R\tau\pi d^2},
\label{eqI}
\end{equation}

\noindent where $R$ is the 1\,kHz repetition rate of the laser. The incident electric field strength $E_i$ is calculated via

\begin{equation}
E_i=\sqrt{\frac{2I}{c\epsilon_0}},
\label{eqE}
\end{equation}

\noindent with the speed of light $c$ and the vacuum permitivity $\epsilon_0$. The ratio between the incoming field $E_i$ and the transmitted field $E_t$ is given by the Fresnel equation for p-polarized light:

\begin{equation}
\frac{E_t}{E_i}=\frac{2\cos{\theta_i}}{n\cos{\theta_i}+\cos{\theta_t}},
\end{equation}

\noindent where $n=3.42$ is the refractive index of the silicon substrate at 6.6\,$\mu$m, and $\theta_i=25^{\circ}$ is the angle of incidence of the light. $\theta_t$ is related to $\theta_i$ via Snell's law. The field amplitude projected into the surface of the sample along the direction of the wires $E_{t,x}$ can be calculated via

\begin{equation}
\frac{E_{t,x}}{E_i}=\frac{E_t\cos{\theta_t}}{E_i}.
\end{equation} 

\noindent All things considered, the field strength in the surface of the sample in the direction parallel to the wires is given by

\begin{equation}
E_{t,x}=0.44E_i.
\end{equation}

\subsection{Laser-assisted photoemission}

For pump photon energies in the mid-infrared spectral range there is a strong coherent interaction between the pump pulse and the photoemitted electrons in a time-resolved ARPES experiment. This ``laser-assisted photoemission'' (LAPE) \cite{SaathoffPhysRevA2008}, where the kinetic energy of the photoeletron is increased or decreased by integer multiples of the pump photon energy, results in the formation of replica bands in the ARPES spectrum. The number of replica bands that is visible in the ARPES spectrum is determined by the pump field strength: the higher the field, the higher the order of replica bands that appear. 

The LAPE effect is particularly strong (absent) when the polarization of the pump pulse is parallel (perpendicular) to the direction in which the photoelectrons are detected. In the present experiment there is a significant component of the pump field along the direction of the detected photoelectrons which means that LAPE dominates the pump-probe signal in the presence of the pump pulse at $t=0$\,ps.

\begin{figure}[h]
	\center
		\includegraphics[width = 1\columnwidth]{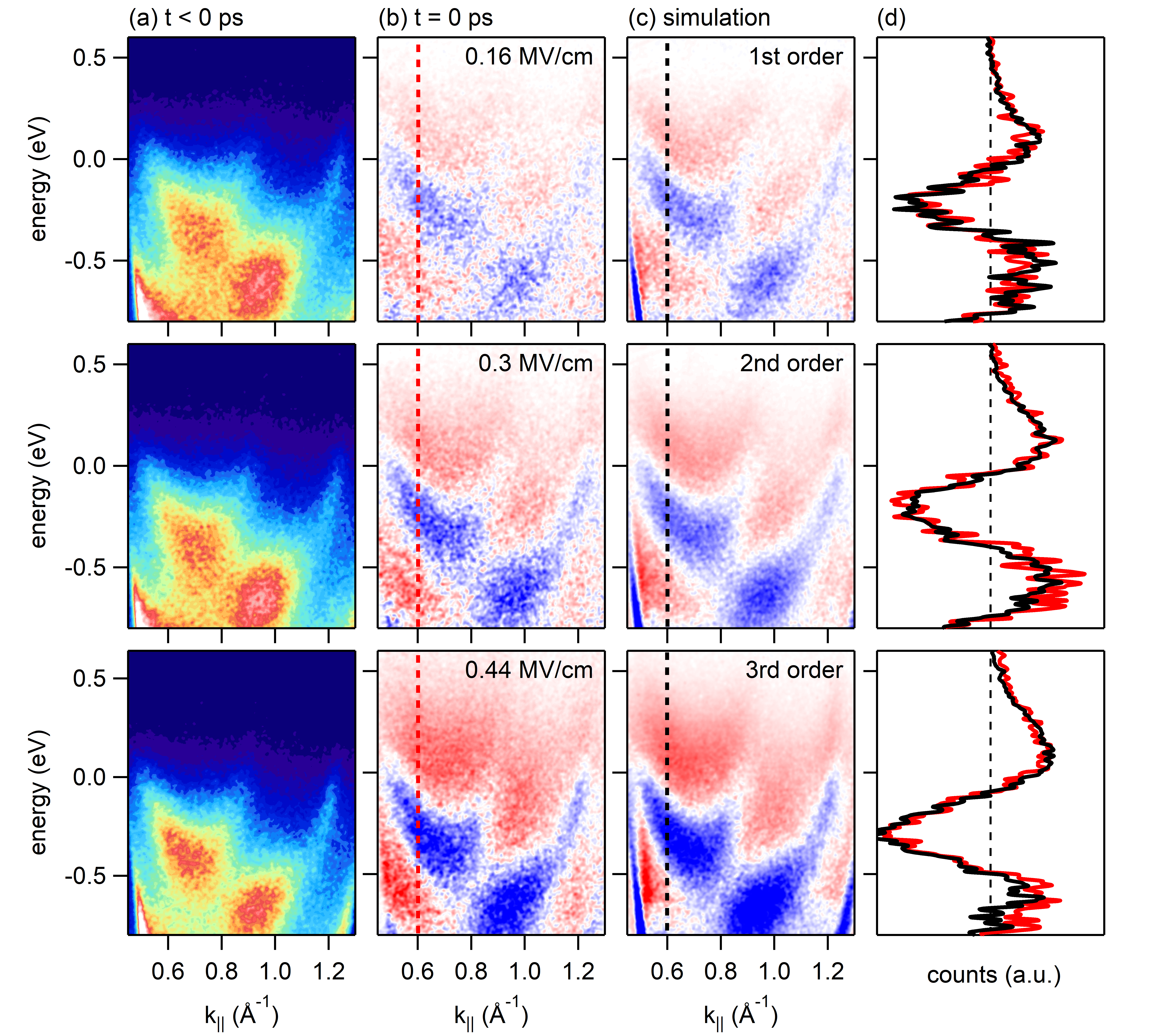}
  \caption{\textbf{Laser-assisted photoemission.} (a) Metallic band structure at negative time delay. (b) Pump-probe signal at $t=0$\,ps with a peak electric field of 0.16\,MV/cm (top), 0.30\,MV/cm (middle), and 0.44\,MV/cm (bottom). (c) Simulated pump-probe signal assuming the generation of first order replica bands (top), first and second order replica bands (middle), and first, second, and third order replica bands (bottom). (d) Direct comparison of measured (red) and simulated (black) energy distribution curves extracted along the dashed line in panels b and c.}
  \label{fig5}
\end{figure}

In Fig. \ref{fig5} we provide evidence that the pump-probe signal at $t=0$\,ps in Fig. 2b of the manuscript is given by the formation of replica bands. Figure \ref{fig5}a shows the photocurrent at negative pump-probe delay at room temperature in the metallic phase. Figure \ref{fig5}b shows the pump-induced changes at $t=0$\,ps for pump pulses at $\hbar\omega=170$\,meV for different field strengths of 0.16\,MV/cm, 0.30\,MV/cm and 0.44\,MV/cm from top to bottom, respectively. In order to simulate the pump-probe signal we take the spectrum measured at negative delay and shift it up and down in energy by integer multiples of the pump photon energy. The simulated differential spectra are shown in Fig. \ref{fig5}c. In Fig. \ref{fig5}d we show selected measured (red) and simulated (black) energy distribution curves to show the excellent quantitative agreement. The experimental data is well reproduced with first order replica bands for a field strength of 0.16\,MV/cm, first and second order replica bands for a field strength of 0.30\,MV/cm, and first, second, and third order replica bands for a field strength of 0.44\,MV/cm.

\subsection{Keldysh parameter}

The Keldysh parameter is given by

\begin{equation}
\gamma=\frac{2\pi\nu\sqrt{2m^{*}E_{\text{gap}}}}{eE},
\end{equation}

\noindent where $\nu$ is the frequency of the pump, $m^{*}$ is the effective mass, $E_{\text{gap}}=300$\,meV is the size of the band gap, $e$ is the charge of the electron, and $E$ is the peak electric field. The effective mass $m^{*}$ is determined from parabolic fits of the valence and conduction band of the $(8\times2)$ structure in \cite{KimPhysRevB2016} as

\begin{equation}
\frac{1}{m^{*}}=\frac{1}{m_{\text{VB}}}+\frac{1}{m_{\text{CB}}}
\end{equation}

\noindent yielding $m^{*}=0.08 m_e$, where $m_e$ is the mass of the free electron. For a field strength of 0.9\,MV/cm we get a Keldysh parameter of $\gamma=1.6$.

\subsection{Model}

\begin{figure}[htp!]
	\centering
	\includegraphics[width=0.5\columnwidth]{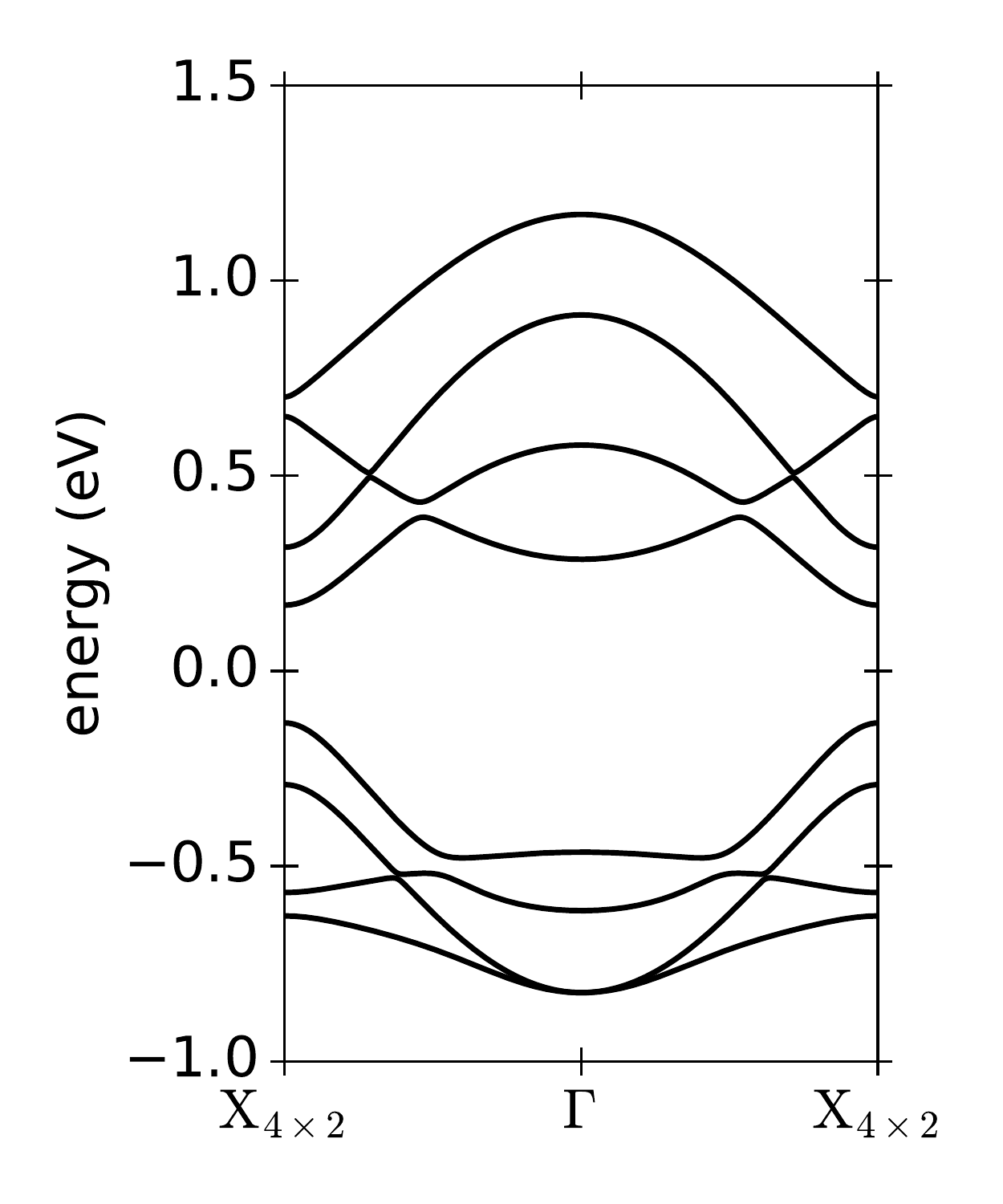}
	\caption{Band structure of the broken symmetry phase with a 300\,meV direct band gap at the $\mathrm{X}$-point of the reduced 4 $\times$ 2 Brillouin zone.}
	\label{fig:bands_model}
\end{figure}

We estimate the critical field strength necessary to melt the CDW by the MIR pump pulse by computing the absorbed energy in a tight-binding model. We use the model Hamiltonian introduced in the Supplementary Material of \cite{JeckelmannPhysRevB2016}, with parameters adjusted to reproduce the reported direct band gap of $\sim$300\,meV. We use the following parameters:	$\epsilon_0 = 0.056$\,eV, $\epsilon_I = 0.089$\,eV (on-site potentials), $t_0 = -0.29$\,eV, $t_0' = -0.545$\,eV,  $t_{I1} = 0.1$\,eV, $t_{I1}' = 0.55$\,eV, $t_{I2} = -0.104$\,eV and $t_{I0} = 0.147$\,eV (next-neighbour hopping integrals). The corresponding band structure in the reduced ($4\times2$) Brillouin zone is shown in Fig. \ref{fig:bands_model}. The key feature relevant for the discussion in the main text is the $\sim$300\,meV direct band gap at the reduced zone boundary.

We set $e=\hbar=c=1$, sample the one-dimensional reduced BZ with 1024 k-points, and set the temperature to $T=40$\,K. The pump pulse is included via a time-dependent vector potential $\bold{A}(t)= A_{\text{max}}p_{\sigma_{p}}(t)\sin(\Omega t)\bold{e}_x$ with a Gaussian envelope $p_{\sigma_{p}}(t)=\text{exp}(-(t-t_0)^2/(2\sigma_p^2))$ and a linear polarization along the wire $\bold{e}_x$ via Peierls substitution in the tight-binding model. We use the mid-point Euler real-time propagation scheme with a time step of $\delta t = 0.07$\,fs. The peak electric field strength is related to the maximum of the vector potential as $E_\text{max} = \Omega A_\text{max}$ \cite{SentefNatComm2015}. 

As in the experiment, we use a pump photon frequency $\Omega=190$\,meV and a pulse duration of 300\,fs at full width at half maximum. The reported absorbed energy (Fig. 4a of the main text) is then computed from the difference of the total energy before and after the pump.


\begin{thebibliography}{12}


\bibitem{CavalleriPhysRevLett2001} A. Cavalleri, C. T{\'o}th, C.W. Siders, J. A. Squier, F. R{\'a}ksi, P. Forget, and J. C. Kieffer, Phys. Rev. Lett. 87, 237401 (2001)
\bibitem{IwaiPhysRevLett2003} S. Iwai, M. Ono, A. Maeda, H. Matsuzaki, H. Kishida, H. Okamoto, and Y. Tokura, Phys. Rev. Lett. 91, 057401 (2003)
\bibitem{CholletScience2005} M. Chollet, L. Guerin, N. Uchida, S. Fukaya, H. Shimoda, T. Ishikawa, K. Matsuda, T. Hasegawa, A. Ota, H. Yamochi, G. Saito, R. Tazaki, S. Adachi, and S. Koshihara, Science 307, 86 (2005)
\bibitem{KusarPhysRevLett2008} P. Kusar, V. V. Kabanov, J. Demsar, T. Mertelj, S. Sugai, and D. Mihailovic, Phys. Rev. Lett. 101, 227001(2008)
\bibitem{PerfettiPhysRevLett2006} L. Perfetti, P. A. Loukakos, M. Lisowski, U. Bovensiepen, H. Berger, S. Biermann, P. S. Cornaglia, A. Georges, and M. Wolf, Phys. Rev. Lett. 97, 067402 (2006)
\bibitem{SchmittScience2008} F. Schmitt, P. S. Kirchmann, U. Bovensiepen, R. G. Moore, L. Rettig, M. Krenz, J.-H. Chu, N. Ru, L. Perfetti, D. H. Lu, M. Wolf, I. R. Fisher, and Z.-X. Shen, Science 321, 1649 (2008)
\bibitem{RohwerNature2011} T. Rohwer, S. Hellmann, M. Wiesenmayer, C. Sohrt, A. Stange, B. Slomski, A. Carr, Y. Liu, L. M. Avila, M. Kall\"ane, S. Mathias, L. Kipp, K. Rossnagel, and M. Bauer, Nature 471, 490 (2011)
\bibitem{PetersenPhysRevLett2011} J. C. Petersen, S. Kaiser, N. Dean, A. Simoncig, H. Y. Liu, A. L. Cavalieri, C. Cacho, I. C. E. Turcu, E. Springate, F. Frassetto, L. Poletto, S. S. Dhesi, H. Berger, and A. Cavalleri, Phys. Rev. Lett. 107, 177402 (2011)
\bibitem{HellmannNatCommun2012} S. Hellmann, T. Rohwer, M. Kall\"ane, K. Hanff, C. Sohrt, A. Stange, A. Carr, M. M. Murnane, H. C. Kapteyn, L. Kipp, M. Bauer, and K. Rossnagel, Nat. Commun. 3, 1069 (2012)
\bibitem{LiuPhysRevB2013} H. Y. Liu, I. Gierz, J. C. Petersen, S. Kaiser, A. Simoncig, A. L. Cavalieri, C. Cacho, I. C. E. Turcu, E. Springate, F. Frassetto, L. Poletto, S. S. Dhesi, Z.-A. Xu, T. Cuk, R. Merlin, and A. Cavalleri, Phys. Rev. B 88, 045104 (2013)
\bibitem{MathiasNatCommun2015} S. Mathias, S. Eich, J. Urbancic, S. Michael, A. V. Carr, S. Emmerich, A. Stange, T. Popmintchev, T. Rohwer, M. Wiesenmayer, A. Ruffing, S. Jakobs, S. Hellmann, P. Matyba, C. Chen, L. Kipp, M. Bauer, H. C. Kapteyn, H. C. Schneider, K. Rossnagel, M. M. Murnane, and M. Aeschlimann, Nat. Commun. 7, 12902 (2015)
\bibitem{RettigNatCommun2015} L. Rettig, R. Cort{\'e}s, J.-H. Chu, I. R. Fisher, F. Schmitt, R. G. Moore, Z.-X. Shen, P. S. Kirchmann, M. Wolf, and U. Bovensiepen, Nat. Commun. 7, 10459 (2015)
\bibitem{CavalleriPhysRevB2004} A. Cavalleri, T. Dekorsy, H. H. W. Chong, J. C. Kieffer, and R. W. Schoenlein, Phys. Rev. B 70, 161102(R) (2004)
\bibitem{TomeljakPhysRevLett2009} A. Tomeljak, H. Sch\"afer, D. St\"adter, M. Beyer, K. Biljakovic, and J. Demsar, Phys. Rev. Lett. 102, 066404 (2009)
\bibitem{HellmannPhysRevLett2010} S. Hellmann, M. Beye, C. Sohrt, T. Rohwer, F. Sorgenfrei, H. Redlin, M. Kall\"ane, M. Marczynski-B\"uhlow, F. Hennies, M. Bauer, A. F\"ohlisch, L. Kipp, W. Wurth, and K. Rossnagel, Phys. Rev. Lett. 105, 187401 (2010)
\bibitem{WallPhysRevLett2012} S. Wall, B. Krenzer, S. Wippermann, S. Sanna, F. Klasing, A. Hanisch-Blicharski, M. Kammler, W. G. Schmidt, and M. Horn-von Hoegen, Phys. Rev. Lett. 109, 186101 (2012)
\bibitem{RiniAppPhysLett2008} M. Rini, Z. Hao, R. W. Schoenlein, C. Giannetti, F. Parmigiani, S. Fourmaux, A. Fujimori, M. Onoda, S. Wall, and A. Cavalleri,  Appl. Phys. Lett. 92, 181904 (2008)
\bibitem{LiuNature2012} M. Liu, H. Y. Hwang, H. Tao, A. C. Strikwerda, K. Fan, G. R. Keiser, A. J. Sternbach, K. G. West, S. Kittiwatanakul, J. Lu, S. A. Wolf, F. G. Omenetto, X. Zhang, K. A. Nelson, and R. D. Averitt, Nature, 487, 345 (2012)
\bibitem{MayerPhysRevB2015} B. Mayer, C. Schmidt, A. Grupp, J. B\"uhler, J. Oelmann, R. E. Marvel, R. F. Haglund, Jr.,
T. Oka, D. Brida, A. Leitenstorfer, and A. Pashkin, Phys. Rev. B 91, 235113 (2015)
\bibitem{GrayPhysRevB2018} A. X. Gray, M. C. Hoffmann, J. Jeong, N. P. Aetukuri, D. Zhu, H.Y. Hwang, N. C. Brandt, H. Wen, A. J. Sternbach, S. Bonetti, A. H. Reid, R. Krukeja, C. Graves, T. Wang, P. Granitzka, Z. Chen, D. J. Higley, T. Chase, E. Jal, E. Abreu, M. K. Liu, T.-C. Weng, D. Sokaras, D. Nordlund, M. Chollet, R. Alonso-Mori, H. Lemke, J. M. Glownia, M. Trigo, Y. Zhu, H. Ohldag, J. W. Freeland, M. G. Samant, J. Berakdar, R. D. Averitt, K. A. Nelson, S. S. P. Parkin, and H. A. D{\"u}r, Phys. Rev B 98, 045104 (2018)  
\bibitem{KaindlJOptSocAmB2000} R. A. Kaindl, M. Wurm, K. Reinmann, P. Hamm, A. H. Weiner, M. Woerner, J. Opt. Soc. Am. B 17, 2086 (2000)
\bibitem{VentalonJOptSocAmB2006} C. Ventalon, J. M. Fraser, J. P. Likforman, D. M. Villeneuve, P. B. Corkum, M. Joffre,  J. Opt. Soc. Am. B 23, 332-340 (2006)
\bibitem{BunkPhysRevB1999} O. Bunk, G. Falkenberg, J. H. Zeysing, L. Lottermoser, R. L. Johnson, M Nielsen, F Berg-Rasmussen, J. Baker, and R. Feidenhans'l, Phys. Rev. B 59, 12228 (1999) 
\bibitem{AbukawaSurfSci1995} T. Abukawa, M. Sasaki, F. Hisamatsu, T. Goto, T. Kinoshita, A. Kakizaki, and S. Kono, Surf. Sci. 325, 33 (1995)
\bibitem{AhnPhysRevLett2004} J. R. Ahn, J. H. Byun, H. Koh, E. Rotenberg, S. D. Kevan, and H. W. Yeom, Phys. Rev. Lett. 93, 106401 (2004)
\bibitem{MorikawaPhysRevB2010} H. Morikawa, C. C. Hwang, and H. W. Yeom, Phys. Rev. B 81, 075401 (2010)
\bibitem{YeomPhysRevLett1999} H. W. Yeom, S. Takeda, E. Rotenberg, I. Matsuda, K. Horikoshi, J. Schaefer, C. M. Lee, S. D. Kevan, T. Ohta, T. Nagao, and S. Hasegawa, Phys. Rev. Lett. 82, 4898 (1999)
\bibitem{SunPhysRevB2008} Y. J. Sun, S. Agario, S. Souma, K. Sugawara, Y. Tago, T. Sato, and T. Takahashi, Phys. Rev. B 77, 125115 (2008)
\bibitem{ChandolaPhysRevLett2009} S. Chandola, K. Hinrichs, M. Gensch, N. Esser, S. Wippermann, W. G. Schmidt, F. Bechstedt, K. Fleischer, and J. F. McGlip, Phys. Rev. Lett 102, 226805 (2009)
\bibitem{TanikawaPhysRevLett2004} T. Tanikawa, I. Matsuda, T. Kanagawa, and S. Hasegawa, Phys. Rev. Lett. 93, 016801 (2004)
\bibitem{KimPhysRevB2016} S.-W. Kim, and J.-H. Cho, Phys. Rev. B 93, 241408(R) (2016)

\bibitem{GonzalezNewJPhys2005} C. Gonz{\'a}lez, J. Ortega, and F. Flores, New J. Phys. 7, 100 (2005)
\bibitem{JeckelmannPhysRevB2016} E. Jeckelmann, S. Sanna, W. G. Schmidt, E. Speiser, and N. Esser, Phys. Rev. B 93, 241407(R) (2016)
\bibitem{KumpfPhysRevLett2000} C. Kumpf, O. Bunk, J. H. Su, M. Nielsen, R. L. Johnson, R. Feidenhans'l, K. Bechgaard, Phys. Rev. Lett. 85, 4916 (2000)
\bibitem{Supplementary} The Supplemental Material contains further information about sample growth, tr-ARPES setup, data analysis and model including references \cite{StevensPhysRevB1993,  LinJApplPhys1998, FrassettoOptExpress2011, SaathoffPhysRevA2008, SentefNatComm2015}.
\bibitem{StevensPhysRevB1993} J. L. Stevens, M. S. Worthington, and I. S. T. Tsong, Phys. Rev. B 47, 1453 (1993)
\bibitem{LinJApplPhys1998} J.-L. Lin, D. Y. Petrovykh, J. Viernow, F. K. Men, D. J. Seo, and F. J. Himpsel, J. Appl. Phys. 84, 255 (1998)
\bibitem{FrassettoOptExpress2011} F. Frassetto, C. Cacho, C. A. Froud, I. C. E. Turcu, P. Villoresi, W. A. Bryan, E. Springate, and L. Poletto, Opt. Express 19, 19169 (2011) 
\bibitem{SaathoffPhysRevA2008} G. Saathoff, L. Miaja-Avila, M. Aeschlimann, M. M. Murnane, and H. C. Kapteyn, Phys. Rev. A 77, 022903 (2008)
\bibitem{SentefNatComm2015} M. A. Sentef, M. Claassen, A. F. Kemper, B. Moritz, T. Oka, J. K. Freericks, and T. Devereaux, Nature Comm. 6, 704 (2015)
\bibitem{FriggeNature2017} T. Frigge, B. Hafke, T. Witte, B. Krenzer, C. Streub\"uhr, A. Samad Syed, V. Mik\v{s}i{\'c} Trontl, I. Avigo, P. Zhou, M. Ligges, D. von der Linde, U. Bovensiepen, M. Horn-von Hoegen, S. Wippermann, A. L{\"u}ke, S. Sanna, U. Gerstmann, and W. G. Schmidt, Nature 544, 207 (2017)
\bibitem{ChavezPhysRevB2018} M. Chavez-Cervantes, R. Krause, S. Aeschlimann, and I. Gierz, Phys. Rev B 97, 201401(R)(2018)
\bibitem{NicholsonArxiv2018} C. W. Nicholson, A. L{\"u}ke, W. G. Schmidt, M. Puppin, L. Rettig, R. Ernstorfer, and M. Wolf, arXiv:1803.11022 (2018)
\bibitem{KeldyshJExpTheorPhys1965} L. Keldysh, J. Exptl. Theoret. Phys. 20, 1307 (1965)
\bibitem{WippermannPhysRevLett2010} S. Wippermann, W. G. Schmidt, Phys. Rev. Lett. 105, 126102 (2010)
\bibitem{YeomPhysRevB2002} H. W. Yeom, K. Horikoshi H. M. Zhang and K. Ono R. I. G. Uhrberg, Phys. Rev. B. 65, 241307 (2002) 

\end{thebibliography}
\end{document}